# Degraded Broadcast Diamond Channels with Non-Causal State Information at the Source

Min Li[1], Osvaldo Simeone[2] and Aylin Yener[1]

[1]Dept. of Electrical Engineering, The Pennsylvania State University, University Park, PA 16802

[2]Dept. of Electrical and Computer Engineering, New Jersey Institute of Technology, University Heights, NJ 07102

*mxl971@psu.edu, osvaldo.simeone@njit.edu, yener@ee.psu.edu*

February 23, 2012

## Abstract

A state-dependent degraded broadcast diamond channel is studied where the source-to-relays cut is modeled with two noiseless, finite-capacity digital links with a degraded broadcasting structure, while the relays-to-destination cut is a general multiple access channel controlled by a random state. It is assumed that the source has non-causal channel state information and the relays have no state information. Under this model, first, the capacity is characterized for the case where the destination has state information, i.e., has access to the state sequence. It is demonstrated that in this case, a joint message and state transmission scheme via binning is optimal. Next, the case where the destination does not have state information, i.e., the case with state information at the source only, is considered. For this scenario, lower and upper bounds on the capacity are derived for the general discrete memoryless model. Achievable rates are then computed for the case in which the relays-to-destination cut is affected by an additive Gaussian state. Numerical results are provided that illuminate the performance advantages that can be accrued by leveraging non-causal state information at the source.

A shorter version of this paper has been submitted to IEEE International Symposium on Information Theory, 2012. The work of M. Li and A. Yener has been supported in part by the National Science Foundation Grants 0721445 and 0964364. The work of O. Simeone has been supported in part by the National Science Foundation Grant 0914899.

I. INTRODUCTION

We consider a communication channel in which the source wishes to communicate to the destination via the help of two parallel relays and there is no direct link between the source and the destination, as shown in Fig. 1. The *first hop*, from the source to the relays, consists of two noiseless digital links of finite capacity: a common link of capacity $C_1$ (bits per channel use) from the source to both relays and a private link of capacity $C_2$ (bits per channel use) from the source to relay 2. The first hop has thus a degraded broadcast channel (BC) structure. The *second hop*, from the relays to the destination, is a general multiple access channel (MAC) controlled by a random state [1]. It is assumed that (*i*) the entire state sequence that affects the MAC is known to the source before transmission, (*ii*) the state is not available at the relays, and (*iii*) it may or may not be known at the destination. We term this channel model as the state-dependent degraded broadcast diamond channel (SD-DBDC) with non-causal channel state information (CSI) at the transmitter (i.e., CSIT) and with or without CSI at the receiver (CSIR).

The motivation to study this channel stems from the downlink of a distributed antenna system, in which a central unit controls two antennas, e.g., two pico-base stations, via backhaul links, for communication to an active user over a wireless channel, see for example [2]. The backhaul communication may be received by both antennas over a wireless broadcast channel modeled by $C_1$, or received by one of antennas via a dedicated optical fiber cable modeled by $C_2$. In such a system, the state may model the fading coefficients for the MAC between the distributed antennas and the user, or an interference signal affecting this MAC. In the first case, the user can typically measure the fading channels of the MAC, thus obtaining CSIR, while the central unit may be informed about such fading channels, e.g., via dedicated feedback links, thus obtaining CSIT. The pico-base stations, serving as the relays, are not expected to decode the feedback signal from the user, due to a design choice or insufficient signal-to-noise ratio, and thus CSI is assumed to be unavailable at the relays. In the latter case of an interfering signal affecting the MAC, the interference signal may be communicated to the central unit via backhaul links from the interfering transmitters, e.g., another central unit, thus obtaining CSIT, while relays and the user are not informed, thus having no CSIR.

*A. Background and Related Work*

The diamond channel, in which a source communicates to two relays via a general broadcast channel and the relays are connected to the destination via a general state-independent MAC, was introduced by Schein and Gallager in [3] and has been widely studied ever since. For the discrete memoryless (DM) diamond channel, several achievability results were established in [3], while for the Gaussian case, it was shown by [4] that partial-decode-and-forward relaying achieves a rate within one bit of the cut-set bound. Despite all the activity, the capacity of this channel in general is open except for some particular instances [5]–[7].

A relevant special case of the diamond channel is obtained when the BC in the first hop is modeled as two orthogonal, noiseless digital links of finite capacity. We refer to this model as orthogonal broadcast diamond channel (OBDC). The OBDC was first studied by Traskov and Kramer in [8], where upper and lower bounds on the capacity of the DM OBDC were derived. Recently, Kang and Liu [9] proposed a single-letter upper bound for the OBDC with a Gaussian MAC and established the capacity for a special subclass of Gaussian OBDCs. The SD-DBDC studied here is related to the OBDC, with the differences that the first hop is modeled as a *degraded* noiseless broadcast channel and that the MAC in the second hop is *state-dependent*.

A comprehensive review of previous work on channels with states can be found in [10], while the discussion here focuses only on work directly related to the present contribution. Consider first a system as in Fig. 1, but with a single relay and with the relay having full knowledge of the message intended for the destination. Note that in this case, the source-to-relay link, unlike the SD-DBDC, only carries state information and not the message. This channel, which can be seen as a point-to-point system with *coded* CSIT, was studied by Heegard and El Gamal in [11] under the assumption of CSIR. Therein, a general lower bound was derived and shown to be tight for some special cases. In [12], Cemal and Steinberg studied the extension of this single-relay setting to the case with two relays, under the assumption that the relays are informed about the two independent messages to be delivered to the destination and that there is full CSIR. This model can be seen as a MAC with coded CSIT. Assuming that the source-to-relays links are modeled as in Fig. 1 with degraded noiseless channels, the capacity region for this model was characterized. Additionally, inner and outer bounds on the capacity region were derived for the case where the source-to-relays cut consists of separate noiseless links. A related work is also that of Permuter et al. [13], which derived the capacity region for a MAC where the encoders, i.e., the relays of Fig. 1, are connected by finite-capacity links to one another, and the MAC channel depends on two correlated state sequences, each known to only one encoder, and there is full CSIR.

We now focus on related studies that assume no CSIR. For the set-up with a single relay and where the relay is informed about the message, i.e., the coded CSIT problem, an upper bound on the capacity was found in reference [14] and proved to be achievable in some special cases. It is noted that, if the relay was informed about both state and message, the optimal strategy would be Gel'fand-Pinsker (GP) encoding [1], which reduces to Dirty Paper Coding (DPC) [15] in the corresponding Gaussian model with an additive state. The state-dependent MAC with various form of CSIT and no CSIR has been studied in [16]–[21]. Assuming non-causal CSIT, the capacity regions for such MAC models are still unknown except the following special instances: the MAC with one informed encoder and degraded messages [18]; the binary MAC with two additive state sequences, each known to one encoder [20]; and the Gaussian MAC with a common state known to both encoders [16] [22, Chapter 7]. Relay channels with state have also been investigated with various type of state information at the nodes, see for example, [23]–[25]. In particular, in reference [24], Zaidi et al. studied a single relay channel with non-causal CSI at the source and proposed various achievable schemes. Capacity results were also identified for some special cases [24].

*B. Contributions*

In this paper, we study the SD-DBDC model illustrated in Fig. 1 with non-causal CSIT and with or without CSIR. Our contributions are summarized as follows:

- For the DM SD-DBDC with non-causal CSIT and CSIR, we find the capacity. The key ingredient of the achievability is a form of binning inspired by [13], whereby the source selects directly the codewords to be transmitted by the relays in such a way as to adapt them to the given realization of the state sequence. It is demonstrated, similar to [13], that such a joint message and state transmission scheme from the source to the relays is optimal and that it generally outperforms a simple scheme whereby the source sends separate message and state descriptions to the relays, see Section III;

- For the DM SD-DBDC with non-causal CSIT and no CSIR, we first derive an upper bound on the capacity and then propose two transmission strategies. The first proposed strategy operates by sending separate message and state descriptions over the digital links to the relays so as to allow each relay to perform GP coding against the quantized state sequence it reconstructs. We refer the scheme to as *GP coding with quantized states* (GP-QS) at the relays. The second scheme, inspired by [24], [26], instead works by having the source first encode the message via GP coding as if the relays had perfect message and state information. Then it sends one common description of the resulting GP sequence to both relays and one refinement description to relay 2. We refer this scheme to as *quantized GP coding* (QGP). The corresponding lower bounds are derived and presented in Section IV-B to IV-C;

- For the case with non-causal CSIT and no CSIR, we also study the Gaussian SD-DBDC with an additive Gaussian state. Achievable rates based on the proposed GP-QS and QGP schemes are evaluated. Numerical results illuminate the merits of non-causal CSIT at the source node and demonstrate the relative performance between the GP-QS and QGP schemes for the Gaussian SD-DBDC, see Section IV-D.

*Notation*: We denote the probability distribution of a random variable $X$ as $p_X(x) = \Pr[X = x]$, or as $p(x)$ when the meaning is clear from the context. Notation $x^i$ represents vector $[x_1, ..., x_i]$. For an integer $L$, the notation $[1:L]$ denotes the set of integers $\{1, ..., L\}$; for a positive real number $l$, the notation $[1:2^l]$ denotes the set of integers $\{1, ..., \lceil 2^l \rceil\}$, where $\lceil \cdot \rceil$ is the ceiling function. $\mathcal{N}(0, \sigma^2)$ denotes a zero-mean Gaussian distribution with variance $\sigma^2$. Finally, $\mathcal{C}(x)$ is defined as $\mathcal{C}(x) = \frac{1}{2}\log_2(1+x)$.

## II. SYSTEM MODEL AND MAIN DEFINITIONS

In this section, we introduce the model studied in this work. Specifically, the SD-DBDC model, depicted in Fig. 1, is denoted by the tuple $(C_1, C_2, \mathcal{X}_1 \times \mathcal{X}_2 \times \mathcal{S}, p(y|x_1, x_2, s), \mathcal{Y})$, where $C_1$ and $C_2$ are the capacities in bits per channel use of the common link from the source to both the relays, and the private link from the source to relay 2, respectively, $\mathcal{X}_1$ and $\mathcal{X}_2$ are the two input alphabets, $\mathcal{S}$ is the state alphabet, $\mathcal{Y}$ is the output alphabet and $p(y|x_1, x_2, s)$ represents the channel probability mass functions (PMFs) describing the MAC between the relays and the destination. The state sequence $s^n$ is generated in an independent and identically distributed

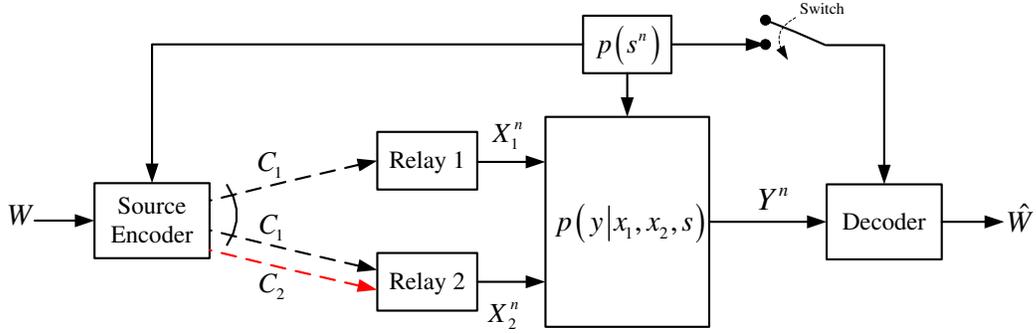

Fig. 1. A state-dependent degraded broadcast diamond channel (SD-DBDC) with non-causal channel state information (CSI) at the transmitter (CSIT) and with or without CSI at the receiver (CSIR). The CSIR switch is closed or open, respectively.

(i.i.d.) fashion according to a fixed PMF $p(s)$, i.e.,

$$p(s^n) = \prod_{i=1}^{n} p(s_i). \qquad (1)$$

The channel is memoryless in the usual sense and the entire state sequence $s^n$ is assumed to be non-causally known to the source node, i.e., we assume non-causal CSIT. However, sequence $s^n$ may or may not be available at the decoder, i.e., we may or may not have CSIR.

Let $W$ be the message that the source wishes to send to the destination, which is uniformly distributed over the set $\mathcal{W} = [1 : 2^{nR}]$. We define the code as follows.

*Definition 1:* A $(2^{nR}, n)$ code for the SD-DBDC consists of:

1) An encoding function at the source node

$$f : \mathcal{W} \times \mathcal{S}^n \to [1 : 2^{nC_1}] \times [1 : 2^{nC_2}], \qquad (2)$$

which maps the message and the state sequence into two indices $M_1$ and $M_2$ transmitted over the source-to-relays links;

2) Two encoding functions at the relays

$$h_1 : [1 : 2^{nC_1}] \to \mathcal{X}_1^n, \qquad (3)$$

$$\text{and } h_2 : [1 : 2^{nC_1}] \times [1 : 2^{nC_2}] \to \mathcal{X}_2^n, \qquad (4)$$

that map the information received by each relay, namely $M_1$ by relay 1 and $(M_1, M_2)$ by relay 2, into the corresponding sequences transmitted by the two relays;

3) A decoding function at the destination. For the case of no CSIR, we have

$$g: \mathcal{Y}^n \to \mathcal{W}, \tag{5}$$

which maps the received sequence into a message estimate $\hat{W} \in \mathcal{W}$, while with CSIR, we have

$$g: \mathcal{Y}^n \times \mathcal{S}^n \to \mathcal{W}, \tag{6}$$

which maps the received sequence and the state sequence into a message estimate $\hat{W} \in \mathcal{W}$.

The average probability of error, $P_e^{(n)}$, is defined as $P_e^{(n)} = \Pr[\hat{W} \neq W]$. A rate $R$ is achievable if there exists a sequence of codes $(2^{nR}, n)$ as defined above such that the probability of error $P_e^{(n)} \to 0$ as $n \to \infty$. The capacity $C$ of this channel is the supremum of the set of all achievable rates [27].

## III. NON-CAUSAL CSIT AND CSIR

In this section, the capacity is established for the DM SD-DBDC with non-causal CSIT and CSIR. The capacity-achieving transmission scheme is presented in Section III-A. For comparison, a straightforward transmission strategy is also considered and its suboptimality is then shown in Section III-B.

### A. Capacity Result

The achievability is based on a scheme in which the source encoder directly selects the codewords to be transmitted by the relays so as to adapt them to the given realization of the state sequence. This is accomplished via a strategy, inspired by [13], in which the codebooks for the transmitted signals $X_1^n$ and $X_2^n$, are binned so that the bin index is identified by the message to be delivered to the destination, and the codewords within the bin are chosen to *match* the state sequence. Moreover, given the degraded broadcast channel between source and relays, the codebooks for $X_1^n$ and $X_2^n$ are superimposed, so that the codeword for $X_1^n$ is known at both relays, while the codeword for $X_2^n$ is only transmitted, superimposed on $X_1^n$, by relay 2. The following theorem presents the result.

*Theorem 1:* For the DM SD-DBDC model with non-causal CSIT and CSIR, the capacity is given by

$$C = \max_{\mathcal{P}} \min \begin{pmatrix} C_1 + C_2 - I(X_1, X_2; S), \\ C_1 - I(X_1; S) + I(X_2; Y | X_1, S), \\ I(X_1, X_2; Y | S) \end{pmatrix} \qquad (7)$$

with the maximum taken over the distributions in the set

$$\mathcal{P} = \{p(s, x_1, x_2, y) : p(s)p(x_1, x_2 | s)p(y | x_1, x_2, s)\} \qquad (8)$$

subject to

$$C_1 \geq I(X_1; S), \qquad (9)$$

$$\text{and } C_1 + C_2 \geq I(X_1, X_2; S). \qquad (10)$$

*Proof:* We provide here a sketch of the proof of achievability. Details are provided in Appendix A, along with the proof of converse. Let $\epsilon_2 > \epsilon_1$, and define functions $\delta(\epsilon_1)$ and $\delta(\epsilon_2)$ such that $\delta(\epsilon_1) \to 0$ as $\epsilon_1 \to 0$ and $\delta(\epsilon_2) \to 0$ as $\epsilon_2 \to 0$. The source splits message $w \in [1 : 2^{nR}]$ into two independent parts $w_1 \in [1 : 2^{nR_1}]$ and $w_2 \in [1 : 2^{nR_2}]$. Message $w_1$ is associated with a bin $\mathcal{B}_1(w_1)$, that contains $2^{n(I(X_1;S)+\delta(\epsilon_1))}$ i.i.d. generated codewords indexed by $x_1^n(w_1, l_1)$, with $l_1 \in [1 : 2^{n(I(X_1;S)+\delta(\epsilon_1))}]$, while message $w_2$ is associated with a bin $\mathcal{B}_2(w_2 | w_1, l_1)$ for all pairs $(w_1, l_1)$, that contains $2^{n(I(X_2;S|X_1)+\delta(\epsilon_2))}$ i.i.d. generated codewords indexed by $x_2^n(w_2, l_2 | w_1, l_1)$, where $l_2 \in [1 : 2^{n(I(X_2;S|X_1)+\delta(\epsilon_2))}]$. Given a message pair $w = (w_1, w_2)$ and a state realization $s^n$, the source encoder first looks for an index $l_1 \in [1 : 2^{n(I(X_1;S)+\delta(\epsilon_1))}]$ such that codeword $x_1^n(w_1, l_1) \in \mathcal{B}_1(w_1)$ is jointly typical with $s^n$; it then looks for an index $l_2 \in [1 : 2^{n(I(X_2;S|X_1)+\delta(\epsilon_2))}]$ such that codeword $x_2^n(w_2, l_2 | w_1, l_1) \in \mathcal{B}_2(w_2 | w_1, l_1)$ is jointly typical with $(x_1^n(w_1, l_1), s^n)$. Thus, index $m_1 = (w_1, l_1)$, is conveyed to both relays and index $m_2 = (w_2, l_2)$, is only conveyed to relay 2 over the digital links. Upon receiving the index and retrieving its corresponding components, relay 1 forwards $x_1^n(w_1, l_1)$ and relay 2 forwards $x_2^n(w_2, l_2 | w_1, l_1)$ to the destination. Observing the output sequence $y^n$ and the state sequence $s^n$, the decoder chooses a unique tuple of $(\hat{w}_1, \hat{w}_2, \hat{l}_1, \hat{l}_2)$ such that $(x_1^n(\hat{w}_1, \hat{l}_1), x_2^n(\hat{w}_2, \hat{l}_2 | \hat{w}_1, \hat{l}_1), s^n, y^n)$ are jointly typical. In this way, the final message estimate $\hat{w}$ is uniquely determined by $\hat{w}_1$ and $\hat{w}_2$. ∎

*B. The Suboptimality of Separate Message-State Transmission*

In the capacity-achieving scheme discussed above, the source encoder selects the codewords for the relay directly based on both message and state sequence in a joint fashion. One can consider, for comparison purposes, a scheme in which the source encoder sends message and state information to the relays separately. The suboptimality of such an approach for a related model was discussed in [13]. We emphasize, however, that, while related, the model considered here is not subsumed by, nor does it subsume, the model in [13].

To elaborate, assume that the source splits the message as $w = (w_1, w_2)$, as done above, and describes the state sequence using a successive refinement code $(S_1, S_2)$ [28], where $S_1$ represents the base state description and $S_2$ represents the refined description. Message $w_1$ and state description $S_1$ are sent to both relays, while message $w_2$ and state description $S_2$ are sent only to relay 2. A coding scheme, similar to that of Theorem 1 of [12], can now be devised in which message $w_1$ is transmitted by using a codebook, conditioned on the description $S_1$, while message $w_2$ is encoded by relay 2, superimposed on the codeword encoding $w_1$ and is conditioned on state descriptions $(S_1, S_2)$. The corresponding achievable rate is characterized as

$$R_{\text{separate}} = \max_{\mathcal{P}'} \min \begin{pmatrix} C_1 + C_2 - I(S_1, S_2; S), \\ C_1 - I(S_1; S) + I(X_2; Y | X_1, S, S_1, S_2), \\ I(X_1, X_2; Y | S, S_1, S_2) \end{pmatrix} \quad (11)$$

with the maximum taken over the distributions in the set

$$\mathcal{P}' = \{p(s, s_1, s_2, x_1, x_2, y) : p(s)p(s_1, s_2 | s)p(x_1 | s_1)p(x_2 | x_1, s_1, s_2)p(y | x_1, x_2, s)\} \quad (12)$$

subject to

$$C_1 \geq I(S_1; S), \quad (13)$$

$$\text{and } C_1 + C_2 \geq I(S_1, S_2; S), \quad (14)$$

where the alphabet size of $S_1$ is bounded as $|\mathcal{S}_1| \leq |\mathcal{S}| + 3$ and the alphabet size of $S_2$ is bounded as $|\mathcal{S}_2| \leq |\mathcal{S}|(|\mathcal{S}| + 3) + 2$, by standard cardinality bounding techniques [22, Appendix C]. Note that the constraints (13) and (14) represent the well-known conditions that allow the construction of a successive refinement code with test channel $p(s_1, s_2 | s)$ [28].

We now show that we have in general $R_{\text{separate}} \leq C$ and that this inequality can be *strict*. In particular, for a fixed $p(s)$ and channel PMF $p(y | x_1, x_2, s)$, considering any PMF in the set

10$\mathcal{P}'$ of (12), we have the following Markov chains: $S - S_1 - X_1$, $S - (S_1, S_2) - (X_1, X_2)$ and $(S_1, S_2) - (S, X_1, X_2) - Y$. Based on these chains, we have the following inequalities

$$C_1 \geq I(S_1; S) \geq I(X_1; S), \tag{15}$$

$$C_1 + C_2 \geq I(S_1, S_2; S) \geq I(X_1, X_2; S), \tag{16}$$

$$I(X_2; Y | X_1, S, S_1, S_2)$$
$$= H(Y | X_1, S, S_1, S_2) - H(Y | X_1, X_2, S, S_1, S_2)$$
$$= H(Y | X_1, S, S_1, S_2) - H(Y | X_1, X_2, S)$$
$$\leq I(X_2; Y | X_1, S), \tag{17}$$

and $I(X_1, X_2; Y | S, S_1, S_2)$
$$= H(Y | S, S_1, S_2) - H(Y | X_1, X_2, S, S_1, S_2)$$
$$= H(Y | S, S_1, S_2) - H(Y | X_1, X_2, S)$$
$$\leq I(X_1, X_2; Y | S), \tag{18}$$

which imply that $R_{\text{separate}} \leq C$. We now show with an example that this inequality can be strict.

For the example, we consider the special case of our model obtained with $C_1 = 0$ and $X_1$ taken as a constant, so that the model reduces to the two-hop line network, consisting of the source, relay 2 and the destination (studied also in [13], see Fig. 2 of [13] if $R_2 = 0$ and $p(y | x_1, x_2, s) = p(y | x_2, s)$). Inspired by the binary example considered in [13] in a slightly different context, we then concentrate on the binary model described by

$$Y = SX_2 \oplus Z, \tag{19}$$

where the state $S \sim \text{Bernoulli}(\frac{1}{2})$, the noise $Z \sim \text{Bernoulli}(p_z)$ with $p_z \triangleq \Pr[Z = 1] \in [0, \frac{1}{2}]$, independent of $S$, and $\oplus$ denotes the modulo-sum operation. We further impose a cost constraint on the binary input $X_2$ at relay 2 as $\frac{1}{n} \sum_{i=1}^{n} \mathbb{E}[X_{2,i}] \leq p_{x_2}$ with $p_{x_2} \in [0, \frac{1}{2}]$, where $\mathbb{E}[.]$ denotes the expectation operation. The capacity of this binary example can be derived from Theorem 1 along with the additional input constraint and is given by

$$C = \max \min \begin{pmatrix} C_2 - H_b(\frac{1}{2}(p_0 + p_1)) + \frac{1}{2} H_b(p_0) + \frac{1}{2} H_b(p_1), \\ \frac{1}{2} H_b(p_1 * p_z) - \frac{1}{2} H_b(p_z) \end{pmatrix}, \tag{20}$$

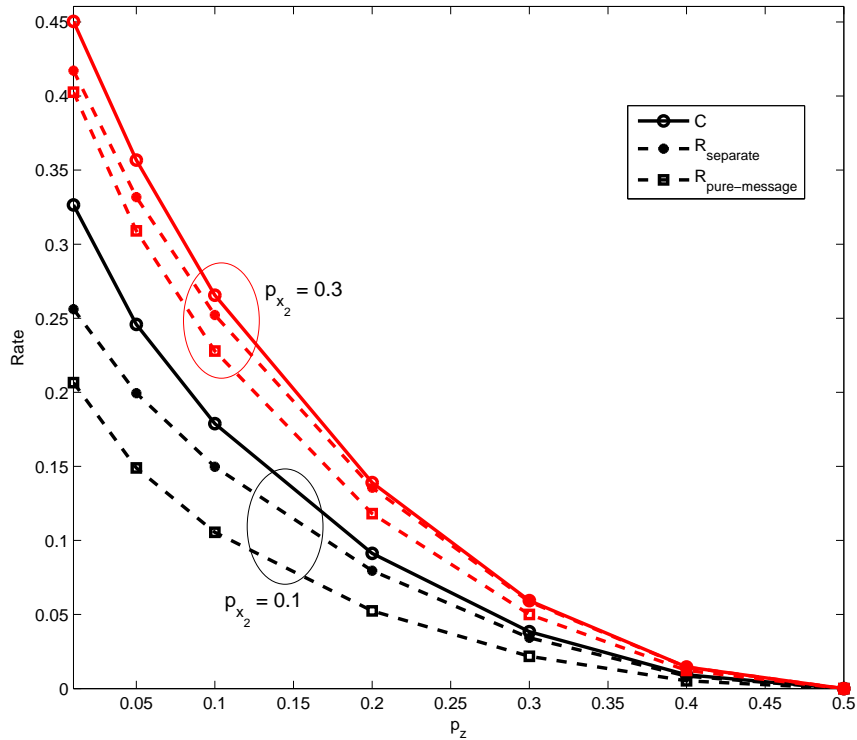

Fig. 2. Performance comparison between $C$, $R_{\text{separate}}$, and $R_{\text{pure-message}}$ for $C_2 = 0.5$, and $p_{x_2} = 0.1$ or $0.3$ in the binary example of Section III-B.

subject to constraints $H_b(\frac{1}{2}(p_0 + p_1)) - \frac{1}{2}H_b(p_0) - \frac{1}{2}H_b(p_1) \leq C_2$ and $\frac{1}{2}(p_0 + p_1) \leq p_{x_2}$, where $p_0 \triangleq \Pr[X_2 = 1 | S = 0] \in [0, 1]$, $p_1 \triangleq \Pr[X_2 = 1 | S = 1] \in [0, 1]$, $H_b(p) \triangleq -p\log_2(p) - (1-p)\log_2(1-p)$, and "$*$" denotes the convolution operation, e.g., $p_1 * p_z = p_1(1 - p_z) + (1 - p_1)p_z$. Similarly, rate $R_{\text{separate}}$ can be obtained from (11). We also consider a special case of the "separate" scheme, in which only message information is sent to the relays, so that we set $S_1$, $S_2$ to a constant in (11) (rate $R_{\text{pure-message}}$ in the figure).

Numerical results are provided in Fig. 2, where $C$, $R_{\text{separate}}$ and $R_{\text{pure-message}}$ are plotted versus $p_z$ for $C_2 = 0.5$, $p_{x_2} = 0.1$ or $0.3$, and the cardinality of $\mathcal{S}_2$ is assumed to be $m = 2$ in $R_{\text{separate}}$ (increasing $m$ to 3, 4 or 5 did not boost the numerical rates of $R_{\text{separate}}$). It is clearly seen that $C$ strictly improves upon $R_{\text{separate}}$ and the latter strictly outperforms $R_{\text{pure-message}}$ for a wide range of $p_z$ values in this example.

IV. NON-CAUSAL CSIT AND NO CSIR

In this section, we turn to the SD-DBDC with non-causal CSIT and without CSIR. In the absence of CSIR, the capacity is difficult to establish. In the following, we thus first present an upper bound on the capacity and then illustrate two achievable schemes for the DM model in Section IV-A to IV-C. Results are then extended to a Gaussian SD-DBDC with an additive Gaussian state in Section IV-D.

*A. An Upper Bound*

*Proposition 1:* For the DM SD-DBDC model with non-causal CSIT and no CSIR, the capacity is upper bounded by

$$R_{\text{upp}} = \max_{\mathcal{P}_{\text{upp}}} \min \begin{pmatrix} C_1 + C_2 - I(X_1, X_2; S), \\ C_1 - I(X_1; S) + I(X_2; Y | X_1, S), \\ I(U; Y) - I(U; S) \end{pmatrix} \qquad (21)$$

with the maximization taken over the distributions in the set

$$\mathcal{P}_{\text{upp}} = \{p(s, u, x_1, x_2, y) : p(s)p(u\,|s)p(x_1, x_2\,|u, s)p(y\,|x_1, x_2, s)\}. \qquad (22)$$

*Proof:* Since the capacity with CSIR cannot be smaller than without CSIR, the first two bounds follows from the converse proof of Theorem 1. The third bound in (21) is instead obtained by providing message and state information to the relays and thus the proof can be derived as in [1] with the identification of auxiliary random variable as $U_i = (W, S_{i+1}^n, Y^{i-1})$. ∎

*B. Achievable Scheme 1: GP Coding With Quantized States At The Relays*

In the absence of CSIR, the source can provide information about the state to the relays so as to allow the latter to perform GP coding. Following this idea and an appropriate combination of message splitting, superposition coding and successive refinement coding [28], similar to the discussion in the previous section, we can devise a scheme detailed below, which is referred to as GP coding with quantized states (GP-QS) at the relays. The GP-QS leads to an achievable rate given as follows.

*Proposition 2:* For the DM SD-DBDC model with non-causal CSIT and no CSIR, a lower bound on the capacity is given by

$$R_{\text{GP-QS}} = \max_{\mathcal{P}_1} \min \begin{pmatrix} C_1 + C_2 - I(S_1, S_2; S), \\ C_1 - I(S_1; S) + I(U_2; Y | U_1) - I(U_2; S_1, S_2 | U_1), \\ I(U_1, U_2; Y) - I(U_1; S_1) - I(U_2; S_1, S_2 | U_1) \end{pmatrix} \quad (23)$$

with the maximum taken over the distributions in the set

$$\mathcal{P}_1 = \left\{ \begin{array}{l} p(s, s_1, s_2, u_1, u_2, x_1, x_2, y): \\ p(s)p(s_1, s_2 | s)p(u_1 | s_1)p(u_2 | u_1, s_1, s_2) \\ p(x_1 | u_1, s_1)p(x_2 | x_1, u_1, u_2, s_1, s_2)p(y | x_1, x_2, s) \end{array} \right\} \quad (24)$$

subject to

$$I(S_1; S) \leq C_1, \quad (25)$$

$$\text{and } I(S_1, S_2; S) \leq C_1 + C_2. \quad (26)$$

*Sketch of Proof:* The proof follows from rather standard arguments, and thus it is only sketched here. Let $\epsilon_2 > \epsilon_1$, and define functions $\delta(\epsilon_1)$ and $\delta(\epsilon_2)$ such that $\delta(\epsilon_1) \to 0$ as $\epsilon_1 \to 0$ and $\delta(\epsilon_2) \to 0$ as $\epsilon_2 \to 0$. As done in the "separate" strategy discussed in the previous section, the source encoder splits message $w \in [1 : 2^{nR}]$ into a common message $w_1 \in [1 : 2^{nR_1}]$, to be delivered to both relays, and a private message $w_2 \in [1 : 2^{nR_2}]$, to be delivered to relay 2 (so that $w = (w_1, w_2)$). Moreover, a successive refinement code $(S_1, S_2)$ is used to describe the state sequence, where the description $S_1$, of rate $R_{s_1}$, is delivered to both relays, and the description $S_2$, of rate $R_{s_2}$, which refines the first, is communicated only to relay 2. As discussed around conditions (13) and (14), the following conditions guarantee the existence of a successive refinement code with test channel $p(s_1, s_2 | s)$

$$R_{s_1} > I(S_1; S), \quad (27)$$

$$\text{and } R_{s_2} > I(S_2; S | S_1). \quad (28)$$

Moreover, in order to guarantee the successful delivery of the messages and state descriptions, the following conditions are sufficient

$$R_1 + R_{s_1} \leq C_1, \quad (29)$$

$$\text{and } R_1 + R_{s_1} + R_2 + R_{s_2} \leq C_1 + C_2. \quad (30)$$

Given the messages and quantized state sequences, GP coding is performed by the relays. Specifically, an auxiliary codebook of $2^{n(R_1+I(U_1;S_1)+\delta(\epsilon_1))}$ i.i.d. codewords $u_1^n$ is generated, and then partitioned into $2^{nR_1}$ bins indexed by $\mathcal{B}_1(w_1)$, where $w_1 \in [1 : 2^{nR_1}]$. Using superposition coding, for each codeword $u_1^n(w_1, l_1)$, where $l_1 \in [1 : 2^{n(I(U_1;S_1)+\delta(\epsilon_1))}]$ is the index of the codeword $u_1^n$ in the bin $\mathcal{B}_1(w_1)$, a second auxiliary codebook of $2^{n(R_2+I(U_2;S_1,S_2|U_1)+\delta(\epsilon_2))}$ i.i.d. codewords $u_2^n(w_2, l_2|w_1, l_1)$ is generated, and then partitioned into $2^{nR_2}$ bins indexed by $\mathcal{B}_2(w_2|w_1, l_1)$, where $w_2 \in [1 : 2^{nR_2}]$ and $l_2 \in [1 : 2^{n(I(U_2;S_1,S_2|U_1)+\delta(\epsilon_2))}]$ is the index of the codeword $u_2^n$ in the bin $\mathcal{B}_2(w_2|w_1, l_1)$. With these codebooks, GP coding of a message $w = (w_1, w_2)$ takes place as follows. Relay 1 and relay 2 encode $w_1$ via the selection of a codeword $u_1^n(w_1, l_1)$ that is jointly typical with the common quantized state sequence $s_1^n$. Then, relay 2 encodes message $w_2$ by choosing a codeword $u_2^n(w_2, l_2|w_1, l_1)$ jointly typical with $(u_1^n(w_1, l_1), s_1^n, s_2^n)$. Appropriate channel inputs $x_1^n$ and $x_2^n$ are then formed by relay 1 and relay 2, respectively, based on the binning codeword(s) selected and the available quantized state(s).

At the destination, upon observing the channel output $y^n$, the decoder looks for a unique pair of $(u_1^n(\hat{w}_1, \hat{l}_1), u_2^n(\hat{w}_2, \hat{l}_2 | \hat{w}_1, \hat{l}_1))$, that is jointly typical with $y^n$, and assigns the message estimate as $\hat{w} = (\hat{w}_1, \hat{w}_2)$. If none or more than one such pair is found, an error is declared. By the packing lemma [22, Chapter 3], it is shown that the probability of decoding error vanishes if

$$R_2 + I(U_2; S_1, S_2 | U_1) < I(U_2; Y | U_1), \tag{31}$$

$$\text{and } R_1 + I(U_1; S_1) + R_2 + I(U_2; S_1, S_2 | U_1) < I(U_1, U_2; Y). \tag{32}$$

Finally, combining the constraints above and using the Fourier-Motzkin procedure [22, Appendix D] to eliminate $(R_{s_1}, R_{s_2})$ and then $(R_1, R_2)$ completes the proof of achievability. ∎

*C. Achievable Scheme 2: Quantized GP Coding*

In the GP-QS scheme, a separate description of state and message is conveyed to the relays. Based on the results with CSIR, one might envision that a scheme in which selection of the relays' codewords is done directly at the source based on both message and state information could be instead advantageous. One such scheme is described here. As further discussed below, however, without CSIR, this scheme is generally not optimal and might even be outperformed by the "separate" GP-QS strategy.

In the second scheme proposed here, inspired by [24], [26], GP coding is done by the source encoder, as if the source encoder had direct access to the relays. Given the finite-capacity link between source and relays, the source encoder then quantizes the resulting GP sequence using a successive refinement code, and conveys a common description to both relays and a private description to relay 2. Upon receiving the descriptions and hence having the reconstructed sequences, the relays simply forward them to the destination. Observing the channel output, the decoder looks for a GP codeword that is jointly typical with the received sequence, and obtains the message estimate as the index of the bin to which such codeword belongs. This scheme is referred to as the quantized GP coding (QGP). It leads to the following achievable rate.

*Proposition 3:* For the DM SD-DBDC model with non-causal CSIT and no CSIR, a lower bound on the capacity is given by

$$R_{\text{QGP}} = \max_{\mathcal{P}_2}(I(U;Y) - I(U;S)) \tag{33}$$

with the maximum taken over the distributions in the set

$$\mathcal{P}_2 = \left\{ \begin{array}{l} p(s,u,v,x_1,x_2,y): \\ p(s)p(u\,|\,s)p(v\,|\,u,s)p(x_1,x_2\,|\,v)p(y\,|\,x_1,x_2,s) \end{array} \right\} \tag{34}$$

subject to

$$I(X_1;V) \leq C_1, \tag{35}$$

$$\text{and } I(X_1,X_2;V) \leq C_1 + C_2. \tag{36}$$

*Remark 1:* The proof of the proposition follows from the discussion above and standard arguments [1], [28] and hence details are omitted for brevity. In the achievable rate derived, we remark that as in [1], $U^n$ denotes the auxiliary binning codewords, while $V^n$ denotes the (auxiliary) analog input sequence, produced by GP encoding at the source encoder. A common description of $V^n$ is carried via both $X_1^n$ and $X_2^n$, a private one is carried via $X_2^n$ only. Inequalities (35)–(36) impose the rates at which the descriptions can be generated. The rate (33) is the rate achievable by GP coding on the virtual channel that connects the source to the destination. □

*Remark 2:* While a general performance comparison between the GP-QS and QGP schemes does not seem to be easy to establish, it can be seen that when the link capacities are arbitrarily large, either the state sequence or the GP analog sequence can be perfectly conveyed to the relays, and thus both the GP-QS and QGP schemes achieve the upper bound (21), and specifically the third bound in (21), thus giving the capacity. □

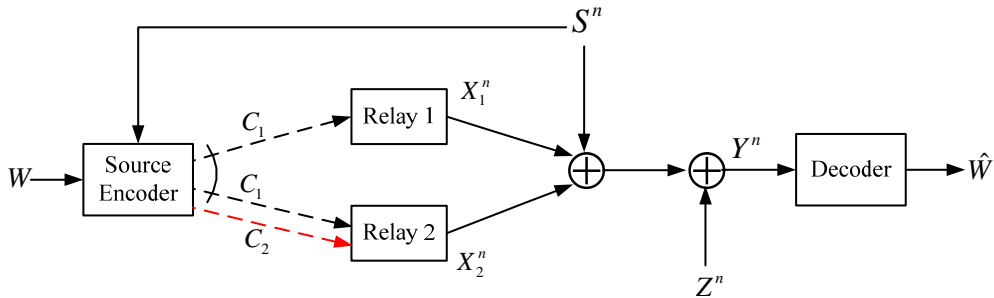

Fig. 3. A Gaussian SD-DBDC with an additive Gaussian state.

*D. Gaussian SD-DBDC*

We now study a Gaussian SD-DBDC as depicted in Fig. 3. In particular, we assume that the destination output $Y_i$ at time instant $i$ is related to the channel inputs $X_{1,i}, X_{2,i}$ at the relays and the channel state $S_i$ as

$$Y_i = X_{1,i} + X_{2,i} + S_i + Z_i, \tag{37}$$

where $S_i \sim \mathcal{N}(0, P_S)$ and $Z_i \sim \mathcal{N}(0, N_0)$, are i.i.d., mutually independent sequences. The channel inputs at the relays satisfy the following average power constraints

$$\frac{1}{n} \sum_{i=1}^{n} \mathbb{E}[X_{k,i}^2] \leq P_k, k = 1, 2. \tag{38}$$

The encoding and decoding functions are defined as in Definition 1 except that the codewords are required to guarantee the input power constraints (38).

*1) Reference Results:* For reference, we first consider the performance of a simple scheme that does not leverage the non-causal CSIT. In particular, the source splits again the message $w$ into two independent parts $w = (w_1, w_2)$ and sends $w_1$ at rate $R_1$ to both the relays and $w_2$ at rate $R_2$ to the relay 2 via the digital links. In this way, the model at hand is converted into a Gaussian MAC channel with degraded message sets [29], [30]. The decoder simply treats the state as noise. The maximum message rates supported by the first hop are given by: $R_1 \leq C_1$ and $R_1 + R_2 \leq C_1 + C_2$, while the capacity region for the MAC cut is obtained from [30] as

$$R_2 \leq \mathcal{C}\left(\frac{(1-\rho^2)P_2}{N_0 + P_S}\right) \tag{39}$$

$$R_1 + R_2 \leq \mathcal{C}\left(\frac{P_1 + P_2 + 2\rho\sqrt{P_1 P_2}}{N_0 + P_S}\right) \tag{40}$$

for $0 \leq \rho \leq 1$, where we recall that $\mathcal{C}(x)$ is defined as $\mathcal{C}(x) = \frac{1}{2}\log_2(1+x)$. Therefore, the overall achievable rate without using CSIT is given by

$$R_{\text{no SI}}^{\text{G}} = \max_{0 \leq \rho \leq 1} \min \begin{pmatrix} C_1 + C_2, \\ \mathcal{C}\left(\frac{P_1 + P_2 + 2\rho\sqrt{P_1 P_2}}{N_0 + P_S}\right), \\ \mathcal{C}\left(\frac{(1-\rho^2)P_2}{N_0 + P_S}\right) + C_1 \end{pmatrix} \tag{41}$$

which serves as a natural lower bound for the capacity of our example considered.

A simple upper bound $R_{\text{upp}}^{\text{G}}$ can be instead obtained by providing the decoder with the interference sequence so that it can be cancelled. The capacity region of the corresponding state-independent system can be found from [30] and is given by (41) with $N_0$ in lieu of $N_0 + P_S$.

*2) Achievable Rates:* We now apply the GP-QS and QGP schemes discussed above to the given Gaussian model.

*Proposition 4:* For the Gaussian SD-DBDC model, the following rate is achievable by the GP-QS scheme:

$$R_{\text{GP-QS}}^{\text{G}} = \max_{\substack{0 \leq \rho \leq 1, \\ (D_1, D_2) \in \mathcal{A}}} \min \begin{pmatrix} C_1 + C_2 - \frac{1}{2}\log_2(\frac{P_S}{D_2}), \\ C_1 - \frac{1}{2}\log_2(\frac{P_S}{D_1}) + \mathcal{C}\left(\frac{\bar{\rho} P_2}{D_2 + N_0}\right), \\ \mathcal{C}\left(\frac{(\sqrt{P_1} + \sqrt{\rho P_2})^2}{\bar{\rho} P_2 + D_1 + N_0}\right) + \mathcal{C}\left(\frac{\bar{\rho} P_2}{D_2 + N_0}\right) \end{pmatrix} \tag{42}$$

where $\bar{\rho} = 1 - \rho$ and the set of $\mathcal{A}$ is defined as

$$\mathcal{A} \triangleq \left\{(D_1, D_2) : P_S \geq D_1 \geq D_2 \geq 0, D_1 \geq P_S 2^{-2C_1}, D_2 \geq P_S 2^{-2(C_1 + C_2)}\right\}. \tag{43}$$

*Proof:* Note that the result of Proposition 2 can be extended to the continuous channel by standard techniques [22, Chapter 3]. Thus, one can obtain the achievable rate in this proposition through evaluation of the general result therein by identifying appropriate inputs. Details of the proof are provided in Appendix B. We remark that $(D_1, D_2)$ in (43) represent the distortions at which the state $S$ is described to the two relays via the successive refinement code $(S_1, S_2)$ used in GP-QS. ∎

Next, we derive the achievable rate based on the QGP scheme.

*Proposition 5:* For the Gaussian SD-DBDC model, the following rate is achievable by the QGP scheme:

$$R_{\text{QGP}}^{\text{G}} = \mathcal{C}\left( \frac{\left(\sqrt{P_1(1-2^{-2C_1})} + \sqrt{P_2(1-2^{-2(C_1+C_2)})}\right)^2}{P_1 2^{-2C_1} + P_2 2^{-2(C_1+C_2)}\left(1 + 2\sqrt{\frac{P_1(1-2^{-2C_1})}{P_2(1-2^{-2(C_1+C_2)})}}\right) + N_0} \right). \qquad (44)$$

*Proof:* The proof is obtained from Proposition 3, similar to the proof of Proposition 4 (see Appendix C). ∎

*Remark 3:* As the digital link capacity $C_1$ becomes arbitrarily large, it is easy to see that both schemes GP-QS and QGP attain the upper bound $R_{\text{upp}}^{\text{G}}$, leading to the capacity $C = \mathcal{C}\left(\frac{P_1+P_2+2\sqrt{P_1 P_2}}{N_0}\right)$. Note that the capacity is the same as if the interference at the destination was not present and if full cooperation was possible at the relays. The benefit of utilizing the non-causal CSIT is therefore evident from this example. We also emphasize that letting capacity $C_2$ alone grow to infinity is not enough to obtain the upper bound above, as in this case only relay 2 can be fully informed by the central unit. □

*Remark 4:* The achievable rate $R_{\text{GP-QS}}^{\text{G}}$ of scheme GP-QS is generally dependent on the interference power $P_S$, while the achievable rate $R_{\text{QGP}}^{\text{G}}$ of scheme QGP is not. This is because in the GP-QS scheme, the state sequence needs to be described to the relays on the finite-capacity links, and thus the stronger is the power $P_S$ of the interfered state, the larger are the feasible distortions $(D_1, D_2)$ in (43) for reproducing the state sequence at the relays. As a result, in the extreme case in which the state power $P_S$ becomes arbitrarily large, the rate $R_{\text{GP-QS}}^{\text{G}}$ reduces to rate $R_{\text{no SI}}^{G}$ (41) obtained when the decoder simply treats the state as noise. On the other hand, in the QGP scheme, the source compresses directly the appropriate GP sequence, whose power does not depend on $P_S$. Given the fact that the performance of QGP is not dependent on $P_S$, it is expected that the QGP scheme outperforms the GP-QS scheme in case $P_S$ is sufficiently large. □

*3) Numerical Results:* We now further investigate the performance of the proposed schemes via numerical results. We first fix the digital link capacities as $C_1 = 1.5$ and $C_2 = 1$. We also set $P_1 = P_2 = P = 1$, and vary $N_0$ so that the signal-to-noise ratio (SNR), defined as $\text{SNR} = 10\log_{10}(P/N_0)$, lies between $[-10 : 30]$ dB. Fig. 4 and Fig. 5 illustrate the corresponding achievable rates versus SNR, given $P_S = 0.2$ or $0.4$, and $P_S = 0.8$ or $1.2$, respectively. It can be seen that with a small state power $P_S$, e.g., $P_S = 0.2$ as in Fig. 4, rate $R_{\text{GP-QS}}^{\text{G}}$ of scheme GP-QS improves upon rate $R_{\text{no SI}}^G$ of the simple scheme without using CSIT, while rate $R_{\text{QGP}}^{\text{G}}$ of scheme QGP is smaller than both. This is due to the fact that, when $P_S$ is relatively small, it is more effective to describe the state sequence to the relays, as done with GP-QS. In the case of moderate $P_S$, e.g., $P_S = 0.4$ as in Fig. 4, we observe that both the GP-QS and QGP schemes outperform the simple scheme. In the case of moderate-to-strong $P_S$, e.g., $P_S = 0.8$ or $1.2$ as in Fig. 5, as explained in Remark 4, scheme QGP is generally advantageous over scheme GP-QS.

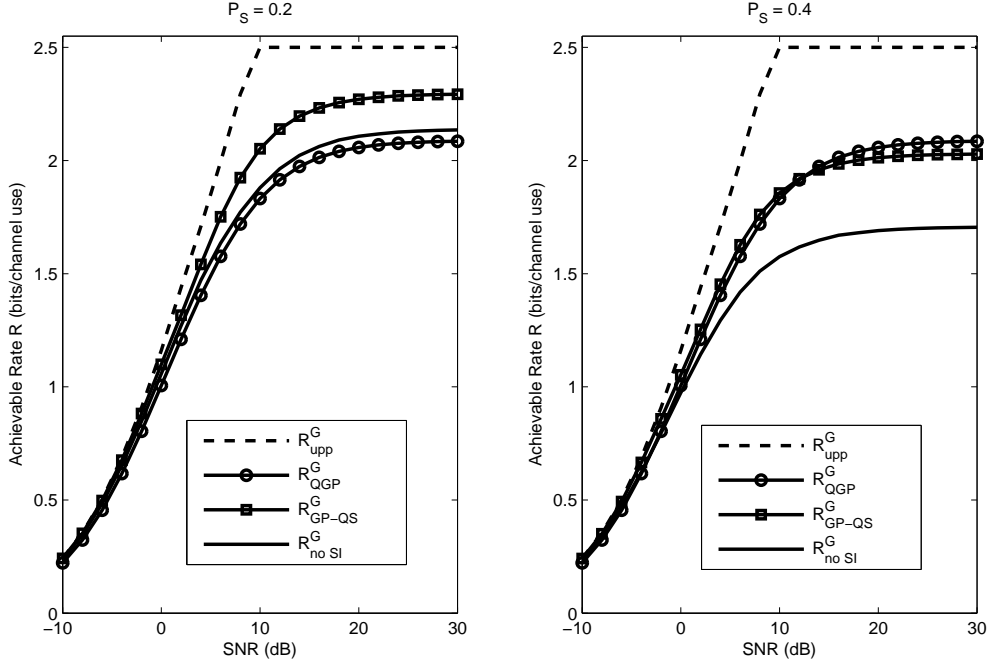

Fig. 4. Achievable rates $R$ vs. SNR for $C_1 = 1.5, C_2 = 1, P_1 = P_2 = 1, P_S = 0.2$ or $0.4$.

We now plot in Fig. 6 the achievable rates versus $C_1$, for $C_2 = 1$, $P_1 = P_2 = 1$, $N_0 = 0.1$ and $P_S = 1.2$. It can be seen that, when $C_1$ is large enough, both the GP-QS and QGP schemes attain the upper bound $R_{\text{upp}}^{\text{G}}$, hence giving the capacity, as discussed in Remark 3. Next, the achievable rates are plotted versus $P_S$ in Fig. 7, for fixed link capacities $C_1 = 1.5, C_2 = 1$ and $P_1 = P_2 = 1$, $N_0 = 0.1$. This figure further confirms the discussion in Remark 4, by showing that both rates $R_{\text{GP-QS}}^{\text{G}}$ and $R_{\text{no SI}}^{\text{G}}$ decrease as $P_S$ increases.

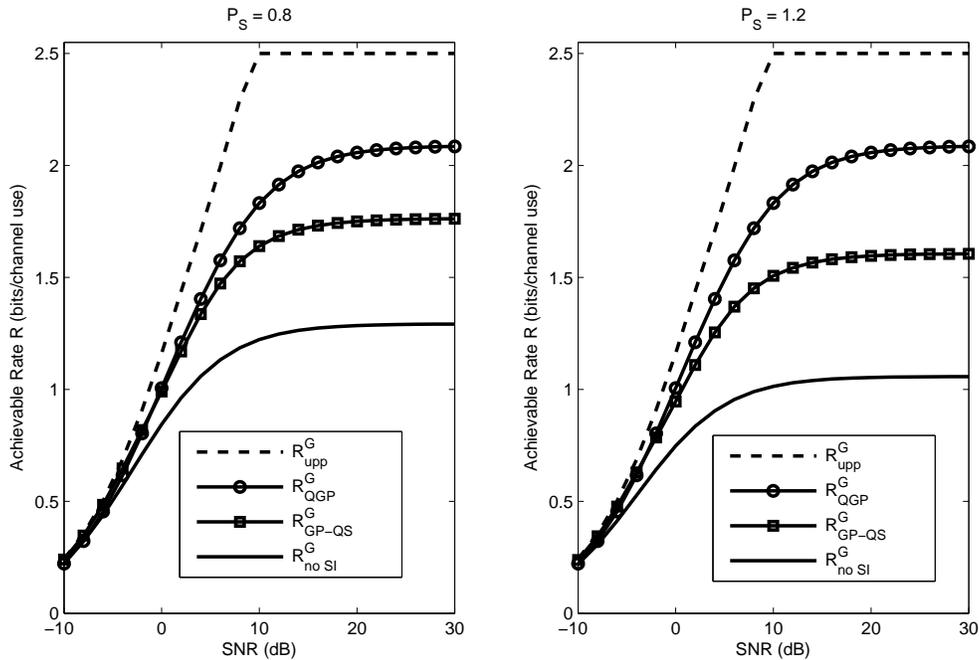

Fig. 5. Achievable rates $R$ vs. SNR for $C_1 = 1.5, C_2 = 1, P_1 = P_2 = 1, P_S = 0.8$ or $1.2$.

## V. CONCLUSIONS

In this work, we have studied a state-dependent diamond channel, in which the broadcast channel between source and relays is defined by a noiseless degraded broadcast channel, and the multiple access channel between relays and destination is state-dependent. For the case with non-causal channel state information at the transmitter (CSIT) and at the receiver (CSIR), we have established the capacity and shown that a joint message and state transmission scheme via binning is optimal and superior to the scheme that performs separate message and state description transmission. For the case without CSIR, we have proposed an upper bound and two transmission schemes, and applied the results to a Gaussian model with an additive Gaussian state. For the Gaussian model, numerical results demonstrate the merit of the non-causal CSIT, and indicate that the best available transmission scheme generally depends on the power of the state. The capacity for the case without CSIR remains open in general and serves as an interesting problem for future work.

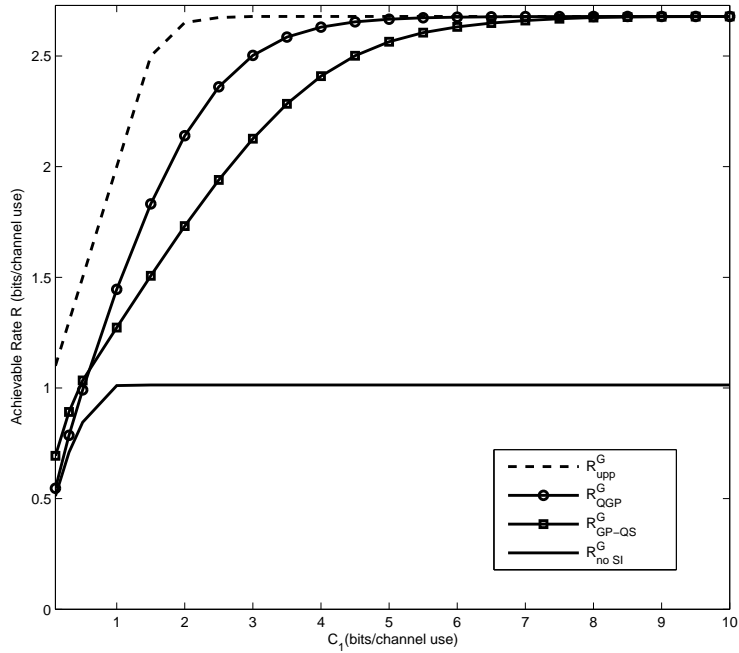

Fig. 6. Achievable rates $R$ vs. $C_1$ for $C_2 = 1, P_1 = P_2 = 1, N_0 = 0.1, P_S = 1.2$.

# APPENDIX A

## PROOF OF THEOREM 1

Throughout the achievability proofs in the paper we use the definition of a strong typical set [22]. In particular, the set of strongly jointly $\epsilon$-typical sequences [22] according to a joint probability distribution $p(xy)$ is denoted by $T_\epsilon^n(XY)$. When the distribution, with respect to which typical sequences are defined, is clear from the context, we will use $T_\epsilon^n$ for short.

**Achievability**:

*Codebook generation*: Fix a joint distribution $p(s)p(x_1, x_2 | s)p(y | x_1, x_2, s)$ where $p(s)$ and $p(y | x_1, x_2, s)$ are defined by the channel. Let $R = R_1 + R_2$, $\tilde{R}_1 > R_1 \geq 0$ and $\tilde{R}_2 > R_2 \geq 0$. Randomly and independently generate $2^{n\tilde{R}_1}$ i.i.d. $x_1^n$ sequences, each according to $\prod_{i=1}^{n} p(x_{1,i})$ and then partition them into $2^{nR_1}$ bins $\mathcal{B}_1(w_1)$, with $w_1 \in \left[1 : 2^{nR_1}\right]$. Hence, there are $2^{n(\tilde{R}_1 - R_1)}$ $x_1^n$ codewords in each bin, which are indexed by $x_1^n(w_1, l_1)$ with $l_1 \in [1 : 2^{n(\tilde{R}_1 - R_1)}]$. Moreover, for any given $x_1^n(w_1, l_1)$, generate $2^{n\tilde{R}_2}$ i.i.d. $x_2^n$ sequences, each according to $\prod_{i=1}^{n} p(x_{2,i} | x_{1,i}(w_1, l_1))$

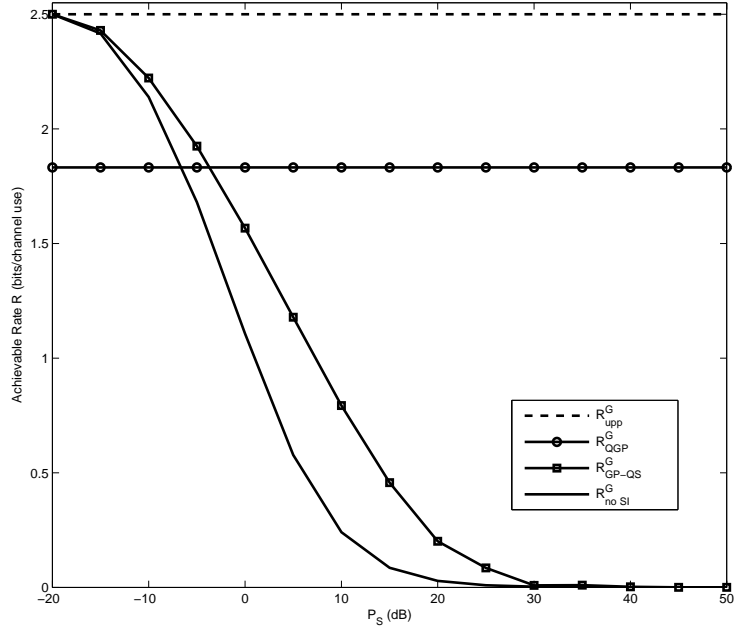

Fig. 7. Achievable rates $R$ vs. $P_S$ for $C_1 = 1.5, C_2 = 1, P_1 = P_2 = 1, N_0 = 0.1$.

and then partition them into $2^{nR_2}$ bins $\mathcal{B}_2(w_2 | w_1, l_1)$, with $w_2 \in [1 : 2^{nR_2}]$. Hence, there are $2^{n(\tilde{R}_2 - R_2)}$ $x_2^n$ codewords in each bin, which are further indexed by $x_2^n(w_2, l_2 | w_1, l_1)$ with $l_2 \in [1 : 2^{n(\tilde{R}_2 - R_2)}]$. Reveal the whole codebook generated to all parties involved.

*Encoding*: Let $\epsilon_3 > \epsilon_2 > \epsilon_1$, and define functions $\delta(\epsilon_k)$ such that $\delta(\epsilon_k) \to 0$ as $\epsilon_k \to 0$ for $k = 1, 2, 3$. The source encoder splits message $w \in [1 : 2^{nR}]$ into two independent parts $w_1 \in [1 : 2^{nR_1}]$ and $w_2 \in [1 : 2^{nR_2}]$. Message $w_1$ is associated with each bin $\mathcal{B}_1(w_1)$, while message $w_2$ is associated with each bin $\mathcal{B}_2(w_2 | w_1, l_1)$ for any fixed $(w_1, l_1)$. Given the message pair $(w_1, w_2)$ and non-causal state information $s^n$, the source encoder first looks for a codeword $x_1^n(w_1, l_1) \in \mathcal{B}_1(w_1)$ such that $(x_1^n(w_1, l_1), s^n) \in T_{\epsilon_1}^n(X_1 S)$; if there are more than one, choose the first one according to the lexicographic order; if there is none, set $l_1 = 1$. Given the $x_1^n(w_1, l_1)$ found, the source encoder further looks for $x_2^n(w_2, l_2 | w_1, l_1) \in \mathcal{B}_2(w_2 | w_1, l_1)$ such that $(x_2^n(w_2, l_2 | w_1, l_1), x_1^n(w_1, l_1), s^n) \in T_{\epsilon_2}^n(X_2 X_1 S)$; if there are more than one, choose the first one according to the lexicographic order; if there is none, set $l_2 = 1$. Then the source conveys index $m_1 = (w_1, l_1)$ and index $m_2 = (w_2, l_2)$ to the relays via the digital links. In particular, index $m_1$ is intended for both relays and $m_2$ only for relay 2. Upon receiving the index and retrieving its corresponding components from the source, relay 1 transmits $x_1^n(w_1, l_1)$, while relay 2 transmits $x_2^n(w_2, l_2 | w_1, l_1)$ to the destination.

*Decoding*: Given $(s^n, y^n)$, the decoder looks for a unique tuple of $(\hat{w}_1, \hat{l}_1, \hat{w}_2, \hat{l}_2)$ such that $(x_1^n(\hat{w}_1, \hat{l}_1), x_2^n(\hat{w}_2, \hat{l}_2 | \hat{w}_1, \hat{l}_1), s^n, y^n) \in T_{\epsilon_3}^n(X_1 X_2 S Y)$; if there is none or more than one such tuples, an error is reported. Then the final message estimate is assigned as $\hat{w} = (\hat{w}_1, \hat{w}_2)$.

*Analysis of probability of error*: Without loss of generality, assume that $w = (w_1, w_2) = (1, 1)$ is sent by the source and the indices conveyed to the relays are $M_1 = (1, L_1)$ and $M_2 = (1, L_2)$. The analysis of probability of error mainly follows from the covering lemma and the packing lemma [22, Chapter 3]. Specifically, by the covering lemma, given any typical sequence $s^n$, the source encoding error vanishes as $n \to \infty$ if

$$\tilde{R}_1 - R_1 > I(X_1; S) + \delta(\epsilon_1), \tag{45}$$

$$\text{and } \tilde{R}_2 - R_2 > I(X_2; S | X_1) + \delta(\epsilon_2). \tag{46}$$

Moreover, the indices $M_1$ and $M_2$ can be perfectly conveyed to both relays and relay 2, respectively, as long as the digital link capacities satisfy

$$\tilde{R}_1 \leq C_1, \tag{47}$$

$$\text{and } \tilde{R}_1 + \tilde{R}_2 \leq C_1 + C_2. \tag{48}$$

By the packing lemma, the probability of decoding error event $\{(w_1, l_1) \neq (1, L_1), \text{for all } (w_2, l_2)\}$ vanishes as $n \to \infty$ if

$$\tilde{R}_1 + \tilde{R}_2 < I(X_1, X_2; Y, S) - \delta(\epsilon_3). \tag{49}$$

Similarly, the probability of decoding error event $\{(w_1, l_1) = (1, L_1), (w_2, l_2) \neq (1, L_2)\}$ vanishes as $n \to \infty$ if

$$\tilde{R}_2 < I(X_2; Y, S | X_1) - \delta(\epsilon_3). \tag{50}$$

Finally, combining the above conditions (45)–(50) and using the Fourier-Motzkin procedure to eliminate $(\tilde{R}_1, \tilde{R}_2)$ and then $(R_1, R_2)$ completes the proof of achievability.

**Converse**: Let $M_1$ be the common index conveyed to both relays and $M_2$ be the private index conveyed to relay 2 only. First, considering the digital link capacity constraint, we have that

$$nC_1 \geq H(M_1) \tag{51}$$

$$\geq I(M_1; S^n) \tag{52}$$

$$= \sum_{i=1}^{n} I(M_1; S_i \,|\, S^{i-1}) \tag{53}$$

$$= \sum_{i=1}^{n} I(M_1, S^{i-1}, X_{1,i}; S_i) \tag{54}$$

$$\geq \sum_{i=1}^{n} I(X_{1,i}; S_i), \tag{55}$$

where (54) holds because of the facts that $S_i$ is independent of $S^{i-1}$ and $X_{1,i}$ is a deterministic function of $M_1$. By the same reasoning, we can show that

$$n(C_1 + C_2) \geq H(M_1, M_2) \tag{56}$$

$$\geq I(M_1, M_2; S^n) \tag{57}$$

$$\geq \sum_{i=1}^{n} I(X_{1,i}, X_{2,i}; S_i). \tag{58}$$

We can also write

$$nR = H(W) \tag{59}$$

$$\leq I(W; Y^n \,|\, S^n) + n\epsilon_n \tag{60}$$

$$= \sum_{i=1}^{n} I(W; Y_i \,|\, Y^{i-1}, S^n) + n\epsilon_n \tag{61}$$

$$= \sum_{i=1}^{n} \left[ H(Y_i \,|\, Y^{i-1}, S^n) - H(Y_i \,|\, Y^{i-1}, S^n, W, M_1, M_2, X_{1,i}, X_{2,i}) \right] + n\epsilon_n \tag{62}$$

$$= \sum_{i=1}^{n} \left[ H(Y_i \,|\, Y^{i-1}, S^n) - H(Y_i \,|\, S_i, X_{1,i}, X_{2,i}) \right] + n\epsilon_n \tag{63}$$

$$\leq \sum_{i=1}^{n} I(X_{1,i}, X_{2,i}; Y_i \,|\, S_i) + n\epsilon_n \tag{64}$$

with $\epsilon_n \to 0$ as $n \to \infty$, where (60) is due to Fano's inequality, i.e., $H(W \,|\, Y^n, S^n) \leq n\epsilon_n$; (62) holds because $(M_1, M_2)$ is a deterministic function of $(W, S^n)$, $X_{1,i}$ is a deterministic function of $M_1$ and $X_{2,i}$ is a deterministic function of $(M_1, M_2)$; (63) follows from the memoryless property of the channel; and (64) follows from the fact that conditioning reduces entropy.

Next, we can prove a second bound on the rate as

$$nR = H(W) \tag{65}$$

$$= H(W\,|S^n) \tag{66}$$

$$= H(W, M_1, M_2\,|S^n) \tag{67}$$

$$= H(M_1, M_2) - I(M_1, M_2; S^n) + H(W\,|M_1, M_2, S^n) \tag{68}$$

$$= H(M_1, M_2) - \sum_{i=1}^{n} I(M_1, M_2, X_{1,i}, X_{2,i}, S^{i-1}; S_i) + H(W\,|M_1, M_2, S^n) \tag{69}$$

$$\leq n(C_1 + C_2) - \sum_{i=1}^{n} I(X_{1,i}, X_{2,i}; S_i) + H(W\,|M_1, M_2, S^n) \tag{70}$$

$$= n(C_1 + C_2) - \sum_{i=1}^{n} I(X_{1,i}, X_{2,i}; S_i) + H(W\,|M_1, M_2, S^n, Y^n) \tag{71}$$

$$\leq n(C_1 + C_2) - \sum_{i=1}^{n} I(X_{1,i}, X_{2,i}; S_i) + n\epsilon_n \tag{72}$$

with $\epsilon_n \to 0$ as $n \to \infty$, where (66) is due to the independence between $W$ and $S^n$; (67) holds because $(M_1, M_2)$ is a deterministic function of $(W, S^n)$; (69) follows from the facts that $S_i$ is independent of $S^{i-1}$, $X_{1,i}$ is a deterministic function of $M_1$ and $X_{2,i}$ is a deterministic function of $(M_1, M_2)$; (70) follows because of the capacity constraints on the links between source and relays, and because of the chain rule and the non-negativity of mutual information; (71) holds due to the Markov chain $W - (M_1, M_2, S^n) - Y^n$ so that $I(W; Y^n\,|M_1, M_2, S^n) = 0$; and (72) follows from Fano's inequality.

Moreover, we have the third bound

$$nR = H(W) \tag{73}$$

$$= H(W, M_1\,|S^n) \tag{74}$$

$$= H(M_1) - I(M_1; S^n) + H(W\,|M_1, S^n) \tag{75}$$

$$\leq nC_1 - \sum_{i=1}^{n} I(X_{1,i}; S_i) + H(W\,|M_1, S^n) \tag{76}$$

$$\leq nC_1 - \sum_{i=1}^{n} I(X_{1,i}; S_i) + I(W; Y^n\,|M_1, S^n) + n\epsilon_n \tag{77}$$

$$= nC_1 - \sum_{i=1}^{n} I(X_{1,i}; S_i)$$
$$+ \sum_{i=1}^{n} \left[ H(Y_i\,|Y^{i-1}, M_1, S^n, X_{1,i}) - H(Y_i\,|Y^{i-1}, M_1, S^n, W, M_2, X_{1,i}, X_{2,i}) \right] + n\epsilon_n \tag{78}$$

$$= nC_1 - \sum_{i=1}^{n} I(X_{1,i}; S_i) + \sum_{i=1}^{n} \left[H(Y_i | Y^{i-1}, M_1, S^n, X_{1,i}) - H(Y_i | S_i, X_{1,i}, X_{2,i})\right] + n\epsilon_n \tag{79}$$

$$\leq nC_1 - \sum_{i=1}^{n} I(X_{1,i}; S_i) + \sum_{i=1}^{n} I(X_{2,i}; Y_i | X_{1,i}, S_i) + n\epsilon_n \tag{80}$$

with $\epsilon_n \to 0$ as $n \to \infty$, where lines (74) to (76) are obtained by similar reasonings for lines (66) to (70) in the previous bound; (77) is due to Fano's inequality, i.e., $H(W | Y^n, S^n, M_1) \leq n\epsilon_n$; (78) holds by the chain rule and also because $M_2$ is a deterministic function of $(W, S^n)$, $X_{1,i}$ is a deterministic function of $M_1$ and $X_{2,i}$ is a deterministic function of $(M_1, M_2)$; (79) follows from the memoryless property of the channel; and (80) holds due to the fact that conditioning reduces entropy.

Finally, let $Q$ be a random variable uniformly distributed over the set $[1:n]$. Define random variables $S = S_Q$, $X_1 = X_{1,Q}$, $X_2 = X_{2,Q}$ and $Y = Y_Q$. Then, bounds (55), (58), (64), (72) and (80) can be written as

$$C_1 \geq I(X_{1,Q}; S_Q | Q) = I(X_1; S | Q), \tag{81}$$

$$C_1 + C_2 \geq I(X_{1,Q}, X_{2,Q}; S_Q | Q) = I(X_1, X_2; S | Q), \tag{82}$$

and

$$R - \epsilon_n \leq I(X_{1,Q}, X_{2,Q}; Y_Q | S_Q, Q) = I(X_1, X_2; Y | S, Q), \tag{83}$$

$$R - \epsilon_n \leq (C_1 + C_2) - I(X_{1,Q}, X_{2,Q}; S_Q | Q) = (C_1 + C_2) - I(X_1, X_2; S | Q), \tag{84}$$

$$R - \epsilon_n \leq C_1 - I(X_{1,Q}; S_Q | Q) + I(X_{2,Q}; Y_Q | S_Q, X_{1,Q}, Q)$$
$$= C_1 - I(X_1; S | Q) + I(X_2; Y | S, X_1, Q), \tag{85}$$

where the distribution on $(Q, S, X_1, X_2, Y)$ from a given code is of the form

$$p(q, s, x_1, x_2, y) = p(q)p(s)p(x_1, x_2 | s, q)p(y | x_1, x_2, s). \tag{86}$$

To eliminate the variable $Q$ from bounds (81) to (85), we note that

$$I(X_1; S | Q)$$
$$= H(S | Q) - H(S | X_1, Q) \tag{87}$$
$$= H(S) - H(S | X_1, Q) \tag{88}$$

$$\geq I(X_1; S), \tag{89}$$

where (88) follows from the fact that the symbols $S_i$ with $i \in [1:n]$ are i.i.d. and hence $S = S_Q$ is independent of $Q$. Similarly, we can prove that

$$I(X_1, X_2; S \mid Q) \geq I(X_1, X_2; S). \tag{90}$$

Moreover, the inequalities

$$I(X_1, X_2; Y \mid S, Q) \leq I(X_1, X_2; Y \mid S), \tag{91}$$

$$\text{and } I(X_2; Y \mid S, X_1, Q) \leq I(X_2; Y \mid S, X_2), \tag{92}$$

hold because of the Markov chain $Q - (X_1, X_2, S) - Y$. Given the facts above, the bounds corresponding to (7)–(10) are recovered by noticing that the distribution of the random variables $(S, X_1, X_2, Y)$ obtained by marginalizing (86) over $Q$ is of the exact form given in $\mathcal{P}$ of (8). This concludes the converse proof and also the proof of Theorem 1.

## APPENDIX B
## PROOF OF PROPOSITION 4

Based on the GP-QS scheme described in Section IV-B and whose achievable rate is given by (23)–(26), for state encoding, we consider the following cascade of backward channels: $S = S_2 + Z_2 = (S_1 + Z_1) + Z_2$, where $S_1 \sim \mathcal{N}(0, P_S - D_1)$, $Z_1 \sim \mathcal{N}(0, D_1 - D_2)$ and $Z_2 \sim \mathcal{N}(0, D_2)$, are independent, and $P_S \geq D_1 \geq D_2 \geq 0$. This construction implies the Markov chain: $S_1 - S_2 - S$. Hence, we have that

$$I(S_1; S) = \frac{1}{2} \log_2 \left( \frac{P_S}{D_1} \right), \tag{93}$$

$$\text{and } I(S_1, S_2; S) = I(S_2; S) = \frac{1}{2} \log_2 \left( \frac{P_S}{D_2} \right). \tag{94}$$

And the constraints of (25) and (26) become

$$D_1 \geq P_S 2^{-2C_1}, \ D_2 \geq P_S 2^{-2(C_1 + C_2)}. \tag{95}$$

For message encoding, we let $X_1 \sim \mathcal{N}(0, P_1)$, independent of $(S, S_1, S_2)$; $X_2 = \sqrt{\frac{\rho P_2}{P_1}} X_1 + V_2$, where $0 \leq \rho \leq 1$, and $V_2 \sim \mathcal{N}(0, \bar{\rho} P_2)$ is also independent of $(S, S_1, S_2)$. The auxiliary random variables $U_1$ and $U_2$ are defined as

$$U_1 = X_1 + \alpha_1 S_1, \tag{96}$$

$$U_2 = V_2 + \alpha_2 \left( S_2 - \alpha_1 \left( 1 + \sqrt{\frac{\rho P_2}{P_1}} \right) S_1 \right), \quad (97)$$

for some $\alpha_1, \alpha_2 \geq 0$ to be specified later. Note that, with these choices, the channel output $Y$ becomes

$$Y = X_1 + X_2 + S + Z \quad (98)$$

$$= \left( 1 + \sqrt{\frac{\rho P_2}{P_1}} \right) X_1 + V_2 + S + Z \quad (99)$$

$$= \left( 1 + \sqrt{\frac{\rho P_2}{P_1}} \right) X_1 + V_2 + S_1 + Z_1 + Z_2 + Z. \quad (100)$$

Therefore, with the choice of $U_1$ given above, we have that

$$I(U_1; Y) - I(U_1; S_1) \leq \mathcal{C} \left( \frac{(\sqrt{P_1} + \sqrt{\rho P_2})^2}{\bar{\rho} P_2 + D_1 + N_0} \right), \quad (101)$$

where the equality is achieved by setting

$$\alpha_1^* = \frac{\left( 1 + \sqrt{\frac{\rho P_2}{P_1}} \right) P_1}{\left( 1 + \sqrt{\frac{\rho P_2}{P_1}} \right)^2 P_1 + \bar{\rho} P_2 + D_1 + N_0} \quad (102)$$

in (96), which is such that $\alpha_1^*(Y - S_1)$ is the minimum Mean-Square-Error (MSE) estimate of $X_1$ given $Y - S_1$, similar to Costa's DPC [15]. Next, to decode the private message carried over $U_2$, the decoder subtracts $\left( 1 + \sqrt{\frac{\rho P_2}{P_1}} \right) U_1$ from $Y$ obtaining the received signal

$$Y' = V_2 + S_2 - \alpha_1^* \left( 1 + \sqrt{\frac{\rho P_2}{P_1}} \right) S_1 + Z_2 + Z. \quad (103)$$

Now, with the choice of $U_2$ in (97), we have that

$$I(U_2; Y | U_1) - I(S_1, S_2; U_2 | U_1) \quad (104)$$

$$= I(U_2; Y') - I \left( U_2; S_2 - \alpha_1^* \left( 1 + \sqrt{\frac{\rho P_2}{P_1}} \right) S_1 \right) \quad (105)$$

$$\leq \mathcal{C} \left( \frac{\bar{\rho} P_2}{D_2 + N_0} \right), \quad (106)$$

where the equality is achieved by setting

$$\alpha_2^* = \frac{\bar{\rho} P_2}{\bar{\rho} P_2 + D_2 + N_0}. \quad (107)$$

This concludes the proof.

APPENDIX C

PROOF OF PROPOSITION 5

Based on the QGP scheme described in Section IV-C and whose achievable rate is given by (33)−(36), we let the auxiliary random variable $V \sim \mathcal{N}(0, P_v)$ for some $P_v > 0$, independent of $S$. Consider the following cascade of forwarding channels: $X_2 = V + Z_2$, and $X_1 = \alpha_1 X_2 + Z_1$, where $X_1 \sim \mathcal{N}(0, P_1)$ and $X_2 \sim \mathcal{N}(0, P_2)$; $Z_1 \sim \mathcal{N}(0, \sigma_1^2)$, $Z_2 \sim \mathcal{N}(0, \sigma_2^2)$, which are independent of each other and also of $V$; Parameters $\alpha_1, \sigma_1^2$ and $\sigma_2^2$ are to be specified. Following this construction, note that $X_1 - X_2 - V$ forms a Markov chain. Therefore, the constraint of (36) becomes $I(X_1, X_2; V) = I(X_2; V) = \frac{1}{2}\log_2\left(\frac{P_2}{\sigma_2^2}\right) \leq C_1 + C_2$. Thus, one can choose $\sigma_2^2 = P_2 2^{-2(C_1+C_2)}$. Then, $P_v = P_2\left(1 - 2^{-2(C_1+C_2)}\right)$ due to the power constraint on $X_2$. Moreover, noting that $\alpha_1^2 P_2 + \sigma_1^2 = P_1$ and $\frac{1}{2}\log_2\left(\frac{P_1}{\alpha_1^2 \sigma_2^2 + \sigma_1^2}\right) \leq C_1$ due to constraint (35), we thus choose $\sigma_1^2 = \frac{P_1 2^{-2C_1}\left(1 - 2^{-2C_2}\right)}{1 - 2^{-2(C_1+C_2)}}$ and $\alpha_1 = \sqrt{\frac{P_1\left(1 - 2^{-2C_1}\right)}{P_2\left(1 - 2^{-2(C_1+C_2)}\right)}}$. The auxiliary random variable $U$ is defined as $U = V + \beta^* S$, where $\beta^*$ is chosen to be the weight of the minimum MSE estimate of $V$ given $Y - S = X_1 + X_2 + Z$, similar to Costa's DPC [15]. In this way, the message rate $R_{\text{QGP}}^{\text{G}} = I(U; Y) - I(U; S) = I(V; X_1 + X_2 + Z)$ which equals (44). This completes the proof.